\documentclass[conference]{IEEEtran}
\usepackage{cite}
\usepackage{amsmath,amssymb,amsfonts}
\usepackage{graphicx}
\usepackage{textcomp}
\usepackage{booktabs}
\usepackage{array}
\usepackage{microtype}
\usepackage{newtxtext,newtxmath}
\usepackage{placeins}
\usepackage{url}
\usepackage[hidelinks]{hyperref}
\usepackage[a4paper,left=20mm,right=20mm,top=29mm,bottom=29mm]{geometry}

\hypersetup{
  pdfauthor={},
  pdftitle={Batch Before You Time: Decision-Scoped Proxy Execution for Timing-Aware Logic Rewriting},
  pdfsubject={Anonymous DATE submission},
  pdfkeywords={logic synthesis, logic rewriting, gate-level simulation, multi-fidelity evaluation}
}
\newcolumntype{L}[1]{>{\raggedright\arraybackslash}p{#1}}

\begin{document}

\title{Batch Before You Time: Decision-Scoped Proxy Execution for Timing-Aware Logic Rewriting}
\author{Pujun Su, Fudan University, pjsu25@m.fudan.edu.cn}
\maketitle

\begin{abstract}
Standard Delay Format (SDF)-annotated switching simulation distinguishes delay-dependent activity among functionally equivalent rewrites, but evaluating every candidate repeats timing, compilation, and replay. A zero-delay proxy can remove timed evaluations, but generating that proxy candidate by candidate can cost more than the timed work it saves. We present Batch Before You Time (BBYT), which compiles all candidates of one rewrite decision into one scoped zero-delay image and either commits a well-separated proxy winner or invokes the unchanged timed chain. Across 12 counterbalanced holdout sequences, BBYT reduces complete candidate-selection time by 18.05\% on average; the design-level reductions are 5.84\% and 30.25\%, with both confidence intervals above zero. On a counterbalanced 8,192-transition C6288 workload, BBYT is 9.24\% faster than the same gate executed with candidate-wise proxy launches. In the five-workload corpus, BBYT removes 32.52\% of timed candidate evaluations and matches exhaustive timed selection on all 250 evaluated decisions. For on-demand timing-aware rewrite selection, proxy execution and fidelity continuation should use the same decision scope.
\end{abstract}

\begin{IEEEkeywords}
logic rewriting, timing-aware evaluation, gate-level simulation, multi-fidelity selection, simulation batching
\end{IEEEkeywords}

\section{Introduction}

Local logic rewriting replaces a Boolean window with one of several mapped, functionally equivalent implementations~\cite{mishchenko2006dag,lin2022novelrewrite}. The Standard Delay Format (SDF) carries implementation timing into analysis and verification~\cite{ieee2001sdf}, while timing-aware gate-level simulation exposes delayed transitions relevant to activity evaluation~\cite{ahmad2014fast,zhang2022gatspi}. In the on-demand selector studied here, applying this chain to every candidate repeats static timing analysis (STA), delay annotation, compilation, and replay on the candidate-selection critical path. This repeated evaluation motivates the complete-boundary optimization studied here.

Multi-fidelity methods allocate queries between cheap biased observations and accurate expensive ones~\cite{poiani2022mf}, while electronic design automation flows use early or learned guidance to prune design points, cuts, and transformations~\cite{ustun2019lamda,neto2021slap,wang2024prunex}. These policies do not make on-demand proxy execution free. In a lazy rewrite flow, candidates and their proxy scores are generated on demand, so proxy compilation and replay lie on the same critical path as timed evaluation. Candidate-wise proxy execution can therefore remove accurate evaluations while increasing complete selection time.

Batch Before You Time (BBYT) makes one rewrite decision the proxy execution unit. It instantiates all equivalent candidates in one scoped zero-delay image, performs one compilation and replay, and recovers one score per candidate. A fixed relative-gap gate then commits the proxy winner or invokes the original exhaustive SDF-annotated chain for the entire decision. Prior simulation systems accelerate kernels or share execution across tests, cycles, or repeated structure~\cite{zhang2022gatspi,zhang2024gl0am,tong2024batchsim,wang2024dry}. BBYT instead shares the mutually exclusive variants of one rewrite decision, recovers one scoped score per variant, and lets those scores control that decision's all-or-none timed continuation; the candidates, objective, gate, and fallback are unchanged.

Across six mirrored sequences on each of two holdouts, all 12 complete-flow effects are positive and average 18.05\%. A controlled three-condition experiment fixes candidates, gate actions, timed work, and selections while decision-scoped packing saves 9.24\% over candidate-wise proxy execution. The same study places candidate-wise execution 17.65~s behind exhaustive timing, whereas BBYT remains 2.38~s ahead. On two natural four-candidate decisions and their fixed prefixes, the packing advantage rises from 0.352~s at two candidates to 1.059~s at four.

This paper makes three contributions:
\begin{itemize}
  \item We introduce \emph{decision-scoped proxy execution} for timing-aware logic rewriting: one equivalence class shares zero-delay compilation and replay while preserving candidate-local scores and the original exhaustive timed fallback.
  \item We derive a complete-cost condition that separates removed timed work from packed proxy and control cost, and show how candidate-wise proxy launches can reverse the expected saving.
  \item We demonstrate an 18.05\% counterbalanced complete-flow reduction, attribute a controlled-workload 9.24\% gain to packing, and match all 250 exhaustive timed selections while removing 32.52\% of timed candidate evaluations.
\end{itemize}

\section{Motivation and Related Work}
\label{sec:motivation}

\subsection{The Complete Candidate-Selection Boundary}

Consider a rewrite decision whose mapped candidates, loads, and input stimulus are available but whose proxy scores have not been computed. Exhaustive timed evaluation runs the accurate chain for every candidate and returns the minimum-energy implementation. A candidate-wise multi-fidelity flow first compiles and replays each candidate at zero delay, applies a gate, and then runs the same timed chain for every candidate in an ambiguous decision. Its complete cost includes the proxy preparation, compilation, replay, parsing, gate, timed continuation, and final write.

This boundary differs from a study that begins with a table of precomputed proxy scores. Precomputation is appropriate when a compatible table already exists, but in our streaming flow, local rewrite candidates are generated only after a window is reached, mapped, and checked for equivalence. Exhaustively materializing every possible proxy score would change the online workflow rather than accelerate its candidate-selection stage. BBYT retains this streaming order: it creates and consumes one decision image before advancing to the next decision.

The gate is beneficial only when removed timed work exceeds proxy and control cost on the complete path. Candidate-wise execution pays one proxy launch per candidate. Decision-scoped execution pays one launch per rewrite decision and exposes all candidate scores together, so its amortization opportunity grows with the number of candidates sharing that decision.

\begin{table}[!t]
\caption{Execution Work for One $k$-Candidate Decision}
\label{tab:modes}
\centering
\footnotesize
\setlength{\tabcolsep}{2.0pt}
\begin{tabular}{L{20mm}ccc}
\toprule
Execution & \shortstack{Zero-delay\\launches} & Gap gate & \shortstack{Timed\\candidates} \\
\midrule
Exhaustive timed & 0 & no & $k$ \\
Candidate-wise proxy & $k$ & yes & 0 or $k$ \\
BBYT & 1 & yes & 0 or $k$ \\
\bottomrule
\end{tabular}
\end{table}

Table~\ref{tab:modes} separates policy from execution. Candidate-wise proxy execution and BBYT use the same all-or-none timed continuation: either no candidate or all $k$ candidates enter the timed chain. Their direct difference is the number of zero-delay images and launches. Exhaustive timed evaluation removes the proxy path but always pays $k$ accurate evaluations. These matched boundaries support two distinct questions: whether the gate-plus-packing flow beats exhaustive timing, and how much of that outcome is caused by packing rather than by the gate.

\subsection{Timing-Aware Rewriting and Multi-Fidelity Selection}

Graph-based rewriting and technology mapping generate alternative implementations while preserving Boolean function~\cite{mishchenko2006dag,lin2022novelrewrite}. Boolean equivalence defines the candidate set; our flow then ranks its members by switching energy computed from timing-aware transition counts. Ahmad and Ciesielski use STA as a block-level timing predictor while retaining dynamic timing simulation at input/output interfaces~\cite{ahmad2014fast}. Their method operates on design blocks and replaces part of SDF back-annotated simulation with prediction. BBYT operates on the candidate set of one rewrite decision: it changes only the launch scope of zero-delay proxy queries and, when the gate is ambiguous, invokes the complete, unchanged SDF chain for every candidate.

Multi-fidelity best-arm methods allocate queries between cheap biased observations and accurate expensive ones~\cite{poiani2022mf}. Design-automation systems also use learned or early-stage information to prune design points, cuts, or transformations, or to guide synthesis-flow search~\cite{ustun2019lamda,neto2021slap,chowdhury2023bullseye,yu2020flowtune,wang2024prunex}. They change what is searched or transformed. BBYT holds the candidate set and fidelity gate fixed and changes how the proxy queries for that one decision are executed.

Published systems accelerate gate-level activity evaluation with graphics processors and test concatenation~\cite{zhang2022gatspi}, optimize zero-delay and re-simulation kernels~\cite{zhang2024gl0am}, batch cycles~\cite{tong2024batchsim}, or share repeated register-transfer-level structure~\cite{wang2024dry}. Their reuse scope lies in the simulation backend, tests, cycles, or common circuit structure. BBYT instead shares mutually exclusive variants within one rewrite decision while preserving a score for each candidate; those scores directly control that same class's all-or-none SDF-annotated continuation.

\section{Decision-Scoped Proxy Execution}
\label{sec:method}

\begin{figure*}[!t]
  \centering
  \includegraphics[width=0.98\textwidth]{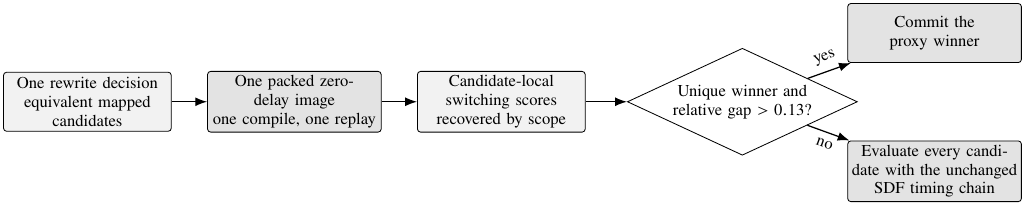}
  \caption{BBYT makes one rewrite decision the shared execution unit. Equivalent candidates share one zero-delay image and replay; candidate-local scores either select a separated proxy winner or invoke the unchanged SDF timing chain for every candidate.}
  \label{fig:method}
\end{figure*}

\subsection{Decision-Level Fidelity Gate}

For decision $d$, let $C_d$ be its set of mapped candidates and let $z_{d,c}$ denote the zero-delay switching energy of candidate $c$. With the proxy values sorted as $z_{d,(1)}\leq z_{d,(2)}\leq\cdots$, BBYT uses the relative gap
\begin{equation}
g_d=\frac{z_{d,(2)}-z_{d,(1)}}{|z_{d,(1)}|}.
\label{eq:gap}
\end{equation}
If the minimum is unique, nonzero, and $g_d>\tau$, the proxy winner is committed. Otherwise every candidate in $C_d$ enters the accurate chain and the minimum timed energy is selected. The threshold is fixed at $\tau=0.13$. Zero minima and proxy ties always use the exhaustive fallback.

For a mapped signal $s$, both fidelities use the same switching-energy calculation
\begin{equation}
E_{d,c}=\frac{1}{2}V^2\sum_{s\in S_{d,c}} C_s N_s,
\label{eq:energy}
\end{equation}
where $C_s$ is its mapped load, $N_s$ is its transition count, and $V$ is the library voltage. The zero-delay and timed fidelities differ only in how $N_s$ is obtained. Thus the gate changes evaluation effort, not the objective used to choose a candidate.

\subsection{One Packed Image per Decision}

Figure~\ref{fig:method} shows the execution path. BBYT renames the candidate top modules without changing their mapped cells or connections, instantiates them under separate scopes in one testbench, and fans out the same stimulus. One compiler invocation creates the image and one replay produces a scoped transition trace. A streaming parser dispatches transitions by candidate scope and applies each candidate's original loads in~\eqref{eq:energy}.

The packed image preserves decision locality. Scores from different rewrite decisions never compete, and only one decision image is live at a time. After scoring, BBYT writes the unique proxy winner or calls the existing static-timing, SDF-generation, compilation, and two-replay path for all candidates in the decision. All 454 direct packed-versus-candidate-wise proxy energies agree, validating the packed naming-and-scoping implementation.

The interface fits a streaming optimizer. The candidate generator hands one fixed set $C_d$, its loads, and its stimulus to the evaluator. BBYT returns either the proxy winner or the winner produced by the existing timed callback before the generator advances. Proxy materialization therefore remains inside the online candidate-selection boundary.

If a decision contains $k=|C_d|$ candidates, candidate-wise proxy execution launches $k$ compilations and replays; BBYT launches one. Image construction is linear in aggregate mapped size, transition demultiplexing is linear in the recorded event stream, and ranking costs $O(k\log k)$. The method therefore trades a larger decision-local image for fewer launches while retaining bounded streaming state.

\subsection{Complete-Cost Break-Even View}

Let $T_{\mathrm{full}}$ and $T_{\mathrm{BBYT}}$ measure the same complete boundary: mapped candidates, loads, and stimulus are ready at the start, and the selected candidate is written at the end. For a fixed workload, their difference can be organized as
\begin{equation}
T_{\mathrm{full}}-T_{\mathrm{BBYT}}
=T_{\mathrm{skipped\ timed}}-T_{\mathrm{packed\ proxy}}-T_{\mathrm{control}},
\label{eq:cost}
\end{equation}
where the first term is the timed candidate work absent from directly committed decisions, the second contains packed proxy materialization, and the last contains selection and residual control work on the method path. BBYT is faster when the removed timed work exceeds both added terms.

Candidate-wise execution uses the same fidelity gate and can remove the same timed candidates, but replaces $T_{\mathrm{packed\ proxy}}$ in~\eqref{eq:cost} with a sum of per-candidate launches. Packing changes no action or fallback work in that comparison. As $k$ grows, the number of avoided launches grows while BBYT still emits one image per decision. Sections~\ref{sec:setup} and~\ref{sec:attribution} test these two implications separately with a same-sequence packing contrast and a natural-candidate multiplicity study.

\begin{table*}[!t]
\caption{Counterbalanced measurements. A complete sequence is the runtime observation.}
\label{tab:experiments}
\centering
\footnotesize
\setlength{\tabcolsep}{2.0pt}
\begin{tabular}{L{22mm}L{26mm}L{50mm}rrL{32mm}}
\toprule
Study & Workload & Balanced conditions & Transitions & Sequences & Sequence contrast \\
\midrule
Holdout confirmation & C6288, EPFL-i2c & BBYT and exhaustive timing; three sequences in each mirrored order per design & 2,048 & 12 & Complete-flow reduction in~\eqref{eq:sequence} \\
Packing attribution & C6288 & Exhaustive, candidate-wise proxy, and BBYT; all six condition orders & 8,192 & 6 & Candidate-wise minus BBYT complete time \\
Candidate multiplicity & Two natural EPFL-cavlc decisions & Candidate-wise and BBYT; mirrored orders at fixed $k=2,3,4$ prefixes & 2,048 & 18 & Candidate-wise minus BBYT proxy component \\
\bottomrule
\end{tabular}
\end{table*}

\section{Experimental Methodology}
\label{sec:setup}

\subsection{Benchmarks, Toolchain, and Configuration}

We evaluate BBYT in a serial open-source timing-aware combinational rewrite flow on 11 circuits from the ISCAS85 and EPFL combinational benchmark suites~\cite{brglez1985iscas,amaru2015epfl}. Six circuits establish implementation consistency; C6288 and EPFL-i2c provide counterbalanced holdout measurements; EPFL-adder, EPFL-max, and EPFL-cavlc extend the five-workload work and quality corpus. Each of these five workloads contains 50 fixed decisions, for 250 decisions and 532 mapped candidates in total.

Yosys 0.67 and its integrated ABC logic-synthesis and verification engine map the candidates and check equivalence~\cite{wolf2013yosys,brayton2010abc}. OpenSTA 3.1 produces delays from the typical Nangate 45-nm library~\cite{cherry_opensta,src2008nangate}, and Icarus Verilog/VVP 14.0 performs zero-delay and SDF-annotated event simulation~\cite{williams_icarus}. Equation~\eqref{eq:energy} uses the library voltage of 1.10~V. Runs use Ubuntu 22.04 on eight logical processors with 7.6~GiB memory; all simulation workloads execute serially.

The relative-gap threshold, candidate sets, and timed objective remain fixed throughout the counterbalanced runs. Primary stimuli contain 2,048 input transitions. The controlled attribution study uses 8,192 transitions generated with the same design seed and preserves the full 2,048-transition prefix. BBYT uses one zero-delay replay; both its fallback and exhaustive timed evaluation use two timed replays per candidate.

\subsection{Counterbalanced Complete-Flow Confirmation}

For each holdout, we run six four-arm sequences. Three place BBYT in arms 1 and 4 with exhaustive timed evaluation in arms 2 and 3; three mirror this order. A sequence effect is the difference between its two exhaustive-arm and two BBYT-arm means, and its reduction divides that difference by the exhaustive mean. The complete sequence, rather than a candidate or an individual arm, is the timing sample. We report two-sided 95\% Student-$t$ intervals over the six sequence effects for each design and a 12-sequence combined sequence summary.

Specifically, if $\overline{T}_{s,\mathrm{BBYT}}$ and $\overline{T}_{s,\mathrm{exh}}$ are the two arm means in sequence $s$, its reported reduction is
\begin{equation}
r_s=1-\frac{\overline{T}_{s,\mathrm{BBYT}}}{\overline{T}_{s,\mathrm{exh}}}.
\label{eq:sequence}
\end{equation}
The design mean and interval are calculated from the six values of $r_s$. Absolute saved-time intervals use the corresponding sequence differences $\overline{T}_{s,\mathrm{exh}}-\overline{T}_{s,\mathrm{BBYT}}$. This preserves the actual repeated unit and avoids treating the decisions inside one process execution as independent runtime observations.

This construction balances each method across the start and end of a sequence and balances the first arm across repeated sequences. Selection identity, candidate count, and number of timed continuations are checked within every sequence; the complete sequence remains the statistical unit.

\subsection{Packing Attribution and Candidate Multiplicity}

The attribution experiment compares three conditions on the same C6288 workload with 8,192 transitions: exhaustive timed evaluation, candidate-wise on-demand proxy execution, and BBYT. We run all six condition orders, so every condition appears twice in each arm position. The main packing contrast is candidate-wise time minus BBYT time within a complete sequence. Both proxy methods use the same threshold, candidates, actions, and timed fallback; a valid contrast additionally requires identical timed work and selected candidates.

The multiplicity experiment uses two naturally occurring four-candidate EPFL-cavlc decisions. Their original candidate-index order defines fixed nested prefixes with $k=2,3,4$; no candidate is copied or reordered. For each $k$, six mirrored four-arm sequences compare candidate-wise and decision-scoped proxy execution. Because timed continuation is unchanged, the primary proxy component subtracts recorded timed-wall contribution from each arm before forming the sequence contrast. The study measures avoided proxy work on the fixed nested prefixes of these two decisions.

All 36 sequences and 138 arms completed. Across these experiments, every comparison records identical actions, timed work, and final selections. Confidence intervals and means in the following sections are computed from the complete sequence table.

\section{Quality, Work, and Complete-Flow Time}
\label{sec:results}

\subsection{Timed Work and Selection Quality}

\begin{table}[!t]
\caption{Fixed-Gate Work and Selection Outcomes}
\label{tab:work}
\centering
\footnotesize
\setlength{\tabcolsep}{1.5pt}
\begin{tabular}{lrrrrr}
\toprule
Workload & \shortstack{Rewrite\\decisions} & \shortstack{Mapped\\candidates} & \shortstack{Timed\\candidates} & \shortstack{Timed\\cut} & \shortstack{Selection\\match} \\
\midrule
C6288       & 50 & 105 & 90 & 14.29\% & 50/50 \\
EPFL-i2c    & 50 & 112 & 64 & 42.86\% & 50/50 \\
EPFL-adder  & 50 & 104 & 86 & 17.31\% & 50/50 \\
EPFL-max    & 50 & 104 & 50 & 51.92\% & 50/50 \\
EPFL-cavlc  & 50 & 107 & 69 & 35.51\% & 50/50 \\
\midrule
Total       & 250 & 532 & 359 & 32.52\% & 250/250 \\
\bottomrule
\end{tabular}
\end{table}

Table~\ref{tab:work} reports the unchanged gate on the five-workload corpus. BBYT runs 359 of 532 candidate-level timed evaluations, removing 173 or 32.52\%. It selects the same mapped candidate as exhaustive timed evaluation on all 250 decisions, so observed timed-energy regret is zero. The per-workload cut ranges from 14.29\% to 51.92\% under one global threshold; decision count alone therefore does not predict the removed work.

The gate commits 81 decisions directly and sends 169 through all-candidate timed fallback. A direct commit removes every timed candidate evaluation in that decision, whereas a fallback retains all of them. The 173 removed candidates therefore arise from the observed candidate multiplicities of the 81 committed decisions rather than from partial candidate pruning.

A post-hoc scan of observed gap breakpoints finds zero selection mismatch from 0.084555 through 0.18, so the fixed 0.13 threshold is not an isolated zero-mismatch breakpoint in this corpus.

\subsection{Counterbalanced Holdout Results}

\begin{table}[!t]
\caption{Counterbalanced Complete-Flow Reduction}
\label{tab:holdout}
\centering
\footnotesize
\setlength{\tabcolsep}{1.2pt}
\begin{tabular}{L{18mm}rrrr}
\toprule
Design & \shortstack{Sequences/\\positive} & \shortstack{Mean\\reduction} & \shortstack{95\%\\interval} & \shortstack{Mean\\saved} \\
\midrule
C6288 & 6/6 & 5.84\% & [3.90, 7.78]\% & 7.66~s \\
EPFL-i2c & 6/6 & 30.25\% & [26.61, 33.90]\% & 42.60~s \\
Combined summary & 12/12 & 18.05\% & [9.78, 26.32]\% & -- \\
\bottomrule
\end{tabular}
\end{table}

\begin{figure}[!t]
  \centering
  \includegraphics[width=\columnwidth]{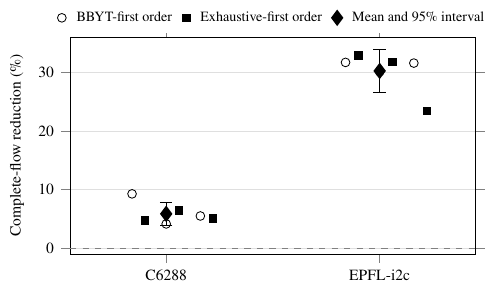}
  \caption{Sequence-level complete-flow reductions. Open circles and filled squares distinguish the two mirrored run orders; diamonds show design means with 95\% confidence intervals. Every sequence is positive.}
  \label{fig:holdout}
\end{figure}

Table~\ref{tab:holdout} and Fig.~\ref{fig:holdout} show positive complete-flow reductions in every sequence on both holdouts. C6288 averages 5.84\% with interval [3.90\%, 7.78\%], and EPFL-i2c averages 30.25\% with interval [26.61\%, 33.90\%]. Both design-level intervals remain above zero.

The absolute sequence effects tell the same story. C6288 saves 7.66~s on average with interval [4.73, 10.59]~s, while EPFL-i2c saves 42.60~s with interval [37.28, 47.91]~s. Across all 12 sequence effects, the combined mean reduction is 18.05\% with interval [9.78\%, 26.32\%]. This combined row summarizes the evaluated sequences; the two design-level intervals remain the primary holdout results.

The BBYT-first and exhaustive-first strata remain positive and have similar means, 18.72\% and 17.38\%, respectively. Thus the measured benefit is present under both mirrored orders rather than being carried by one fixed first arm. C6288 removes 15 timed candidates and EPFL-i2c removes 48, explaining the larger skipped-work term and absolute saving on EPFL-i2c.

\section{Attribution and Candidate Multiplicity}
\label{sec:attribution}

\begin{figure*}[!t]
  \centering
  \includegraphics[width=0.94\textwidth]{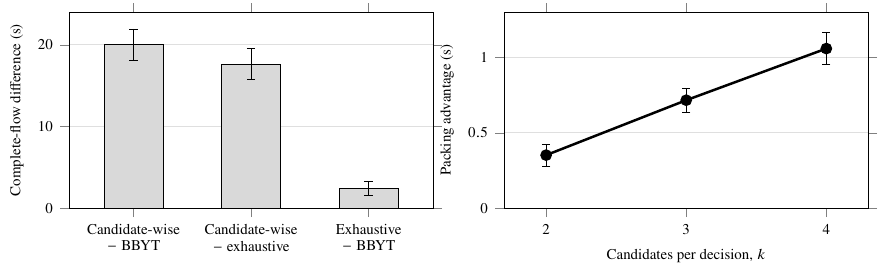}
  \caption{Attribution and multiplicity. (a) Same-sequence C6288 contrasts at 8,192 transitions; each bar is the first condition minus the second, so a positive value means the first takes longer. Error bars are 95\% confidence intervals over six complete sequences. (b) Candidate-wise minus BBYT proxy time for two natural four-candidate decisions and their fixed prefixes; the sequence-level packing advantage increases from $k=2$ to $k=4$.}
  \label{fig:attribution}
\end{figure*}

\subsection{Direct Packing Increment}

Figure~\ref{fig:attribution}(a) isolates proxy execution scope while holding the selection work fixed. Candidate-wise proxy execution takes 20.02~s longer than BBYT on average, with interval [18.10, 21.95]~s; expressed relative to the candidate-wise complete-flow time, decision-scoped packing saves 9.24\%, with interval [8.39\%, 10.10\%]. All six sequence contrasts are positive.

Within the same counterbalanced blocks, candidate-wise proxy execution is 17.65~s slower than exhaustive timed evaluation, with interval [15.73, 19.57]~s, whereas BBYT is 2.38~s faster, with interval [1.50, 3.25]~s. The same proxy signal and gate therefore lose at candidate-wise launch granularity but pass break-even with one image per decision.

Because the two proxy conditions have identical actions, timed work, and selections in all 300 sequence-decision comparisons, their 20.02-s difference identifies the incremental effect of proxy execution scope on this workload.

The three paired contrasts provide a complete accounting of the observed inversion. Candidate-wise proxy execution adds 17.65~s relative to exhaustive timing, while BBYT removes 2.38~s relative to that same reference. Up to reported rounding, these two effects compose the 20.02-s candidate-wise-minus-BBYT contrast. The gain therefore comes from changing how the fixed proxy work is launched, not from changing which decisions continue to timing.

\subsection{Natural Candidate Multiplicity}

Figure~\ref{fig:attribution}(b) tests the amortization implication of decision-scoped execution. On the two evaluated natural four-candidate decisions, candidate-wise minus BBYT proxy time is 0.352~s at $k=2$, 0.716~s at $k=3$, and 1.059~s at $k=4$. The respective 95\% intervals are [0.279, 0.425]~s, [0.635, 0.797]~s, and [0.951, 1.166]~s. All six sequence effects at every prefix are positive.

The complete-wall contrasts move in the same direction: 0.329, 0.665, and 1.023~s for $k=2,3,4$. Their intervals are [0.114, 0.543]~s, [0.530, 0.800]~s, and [0.768, 1.279]~s, respectively. On these two natural four-candidate decisions and their fixed prefixes, the measured packing advantage increases monotonically from $k=2$ to $k=4$ while actions, timed work, and selections remain unchanged in all 18 sequences.

\subsection{System Design Implication}

The experiments separate the terms in~\eqref{eq:cost}: the gate determines removed timed work, decision-scoped execution determines added proxy cost, and candidate multiplicity determines the number of avoided launches. These are distinct design levers. Table~\ref{tab:work} quantifies the gate's effect on timed candidate count, whereas Fig.~\ref{fig:attribution}(a) changes only the execution of proxy work. The 32.52\% timed-candidate reduction therefore describes fidelity allocation, while the 9.24\% paired contrast identifies the additional contribution of packing.

This separation explains why candidate counts alone cannot predict complete-flow time. The candidate-wise condition removes the same timed work as BBYT but remains 17.65~s slower than exhaustive evaluation on the controlled workload. One decision image eliminates enough repeated proxy launches to recover that deficit and finish 2.38~s ahead. A system evaluation that starts after proxy scores have been generated would observe the first lever but omit the cost that determines whether the complete path crosses break-even.

BBYT requires no change to the timed backend. The evaluator receives the candidates, loads, and stimulus for one decision, generates their scoped scores together, and either returns the proxy winner or invokes the existing all-candidate timed callback. This interface preserves the optimizer's streaming order and bounds live state to one decision. It yields more margin when the gate removes more timed work, as on EPFL-i2c, and a larger packing increment when more candidates share a decision.

The relevant execution regime is on-demand proxy generation with repeated compilation and replay. When compatible proxy scores already exist at no marginal cost, the selection stage can consume them directly. When they are generated on the optimization path, the equivalence class is the natural unit for amortizing launch cost because its candidate-local scores jointly decide one timed continuation. The multiplicity evidence covers two natural four-candidate decisions from EPFL-cavlc and their fixed prefixes; it shows the mechanism across $k=2,3,4$ within that evaluated scope.

\section{Conclusion}

BBYT uses one zero-delay image per timing-aware rewrite decision and reserves the unchanged SDF-annotated chain for proxy-ambiguous decisions. Across 12 counterbalanced holdout sequences, it reduces complete candidate-selection time by 18.05\% on average, reproduces all 250 exhaustive timed selections, and removes 32.52\% of timed candidate evaluations. On the controlled 8,192-transition workload, decision-scoped packing contributes 9.24\% relative to candidate-wise proxy execution. On two natural four-candidate decisions from one design and their fixed prefixes, the packing advantage rises monotonically from $k=2$ to $k=4$. For on-demand simulation-in-the-loop rewriting, proxy scores should be generated at the same decision scope that controls timed continuation.
\bibliographystyle{IEEEtran}
\bibliography{references_v4}

\end{document}